 \documentclass[11pt]{article}
 \usepackage[top=1in,bottom=1in,left=1.5in,right=1.5in]{geometry}
 \usepackage{amsmath, amssymb, amsfonts, enumerate}

  \newcommand{\hs}{\hspace*{\parindent}}

\begin{document}

 \title{the Dimer Gas Mayer Series, the Monomer-Dimer $\lambda_d(p)$, the Federbush Relation}

 \author
 {Paul Federbush \\
 Department of Mathematics\\
 University of Michigan \\
 Ann Arbor, MI 48109-1043 \\
 (\texttt{pfed@umich.edu})}
 
  \date{\today}
 \maketitle

 \begin{abstract}
The author and Shmuel Friedland recently presented an expression for $\lambda_d(p)$ of the monomer-dimer problem involving a power series in $p$ . Herein I present a simple way to derive this expression for $\lambda_d(p)$ from the Mayer (or Virial) series of a dimer gas. The derivation is an exercise in basic statistical mechanics. We expect the relationship to be very useful.
 \end{abstract}

Shmuel Friedland and the author recently presented an expression for  $\lambda_d (p)$  of the monomer-dimer problem involving a power series in $p$, see eq (7.1) of \cite{R1}. We always knew that the Mayer or Virial series of a dimer gas implicitly contained all the information to compute our series. (In \cite{R2}, see the remark there in the paragraph following the paragraph containing eq. (30), the sentence beginning: "By a simple device...".) But it was not until the referee of a recent paper, \cite{R3}, asked about the relationship of the expansion for  $\lambda_d(p)$ to the Virial expansion that I finally worked out the details of the simple natural connection. We present the general development, and then specialize to $d=2$ where the most is now known.

We take the statistical mechanical notation from Ruelle, \cite{R4}. We must be careful to avoid confusions, for example between the $p$ in $\lambda_d(p)$ and pressure. We at the outset specify some notation.
\begin{enumerate}[ 1)]
\item $p$ (as in  $\lambda_d(p)$) is considered the \underline{density} of vertices covered by dimers.

\item $\rho$ is the \underline{density} of dimers. So

\begin{equation}
p=2\rho.
\end{equation}

\item $P$ is \underline{pressure}.

\item The \underline{Mayer series} from \cite{R4}, page 84 are 

\begin{equation} 
\beta P=\sum_{n=1}^\infty b_n z^n \label{eq:2}
\end{equation}

and

\begin{equation} 
\rho=\sum_{n=1}^\infty n b_n z^n. \label{eq:3}
\end{equation}

We will set $\beta = 1$. But we choose to work with the density $p$ so we replace (\ref{eq:3}) by
\begin{equation}
p=2\sum_1^\infty n b_n z^n. \label{eq:4}
\end{equation}

\item We solve (\ref{eq:4}) for $z=z(p)$ by iterating (from $z=0$)
\begin{equation}
z=\frac{1}{2b_1}p-\sum_2^\infty \frac{nb_n}{b_1}z^n.
\end{equation}

We then substitute $z=z(p)$ into the right side of (\ref{eq:2}), getting
\begin{equation}
P(p) = \sum_1^\infty b_n(z(p))^n
\end{equation}
Written as a power series in $p$ this is the \underline{Virial series}. 
\end{enumerate}

 Considering the partition function leading to the Mayer series, we note that in volume $V$ the sum is dominated by terms with $\sim \rho V$ dimers, leading to the observation that
\begin{equation}
\lambda_d(p) = P(p) - \frac p 2 \ln(z(p))
\end{equation}
This is the \underline{Federbush relation}. (ADDED HISTORICAL NOTE: Our eq. 7 appears in eq. 6.16 of \cite{R6} and implicitly in \cite{R7}, but not associated to the expansions in p of eq. 6 and eq. 7. The expansion in p of $ \lambda_d(p)$ is what we desire and understand in our eq. 7.) We here assume that the expansion of the left side of this relation given in eq. (7.1) of \cite{R1} is consistent with this relation, 'as it certainly must be'. We leave the proof of this, 'which is just algebra', to another day. Later we will see that for $d = 2$ we have checked this to the $p^7$ power.

We look at $z(p)$ as derived above
\begin{equation}
z=\frac{p}{2b_1}(1+F(p)) \label{eq:7}
\end{equation}
where $F(p)$ is a power series in $p$ without a constant term.We use (8) to get 
\begin{align}
\lambda_d(p) &= P(p) - \frac p 2 \ln\left(\frac{p}{2b_1}\right) - \frac p 2 \ln(1+F(p))\\
&= - \frac p2 \ln(p) + \frac p2 \ln(2d) + P(p) - \frac p2\ln(1+F(p))
\end{align}
We have used that $b_1=d$. Aside from the first two terms with $\ln$'s, the last two terms in (10) are polynomial series in $p$. 

For $d=2$ we calculated as the principal part of our computer computations (in calculating the $\bar J_i$ defined in \cite{R1}) the values of the first 7 coefficients in the Mayer series of a dimer gas:
\begin{align}
b_1&=2\\
b_2&=-7\\
b_3&= \frac{116}{3}\\
b_4&= -\frac{521}{2}\\
b_5&= \frac{9812}{5}\\
b_6&= -\frac{47644}{3}\\
b_7&= \frac{945688}{7}
\end{align}

Equation (10) with the appropriate substitutions yields the same series as in \cite{R3}, up to order $p^7$ in the power series. The computations are easy using Maple:

\begin{align} 
\lambda_2(p) \sim & \frac{1}{2} (p \ln (4) - p \ln p - 2(1-p) \ln (1-p) -p) + \frac{4}{2} \cdot \left( \frac{1}{2 \cdot 1} \left(\frac{p}{4}\right)^2 \right.  \nonumber \\
&\left. + \frac{1}{3 \cdot 2} \left(\frac{p}{4} \right)^3 + \frac{7}{4 \cdot 3} \left( \frac{p}{4} \right)^3 + \frac{41}{5 \cdot 4} \left( \frac{p}{4} \right)^5 + \frac{181}{6 \cdot 5} \left( \frac{p}{4} \right) ^6 + \frac{757}{7 \cdot 6} \left( \frac{p}{4} \right) ^7 \right)
\end{align}

The current development should make it straightforward to prove that the expansion for $\lambda_d(p)$, which is given in (10), converges to the correct value for small enough $p$. We had previously proved in \cite{R5} that the series converged, but not necessarily to the correct value. But we do not see how to recover from the results of the current paper the detailed knowledge of the dependence on dimension $d$ of the earlier derivation.

\end{document}